\documentclass[11pt]{article}

\newcommand\nc{\newcommand}
\nc{\protocolfont}[1]{\textsc{#1}}
\nc\Shelby{\protocolfont{Shelby}\xspace}

\def\thetitle{Rationally Analyzing \Shelby:\\Proving Incentive Compatibility in a Decentralized Storage Network}

\title{\thetitle
        \thanks{We thank Pranav Garimidi, Guru Vamsi Policharla, Tim Roughgarden, and many seminar participants at a16z crypto for helpful comments.
				Goren is employed by Aptos Labs and is a part of the team working on the \Shelby protocol.
				Kominers is a Research Partner at a16z crypto, and part of this work was conducted while Crystal was a Research Intern at a16z crypto; a16z crypto reviewed a draft of this article for compliance prior to publication and is an investor in various crypto projects, including Aptos (for general a16z disclosures, see https://www.a16z.com/disclosures/). Notwithstanding, the ideas and opinions expressed herein are those of the authors, rather than of a16z or its affiliates. 
				Kominers also holds digital assets, including both fungible and non-fungible tokens, and advises a number of companies on marketplace and incentive design, and serves as an expert on related matters. 
}}

\author{Michael Crystal\thanks{Stanford University} \and Guy Goren\thanks{Aptos Labs} \and Scott Duke Kominers\thanks{Harvard University and a16z crypto}}
	
\date{\today}

\usepackage[utf8]{inputenc}
\usepackage[T1]{fontenc}
\usepackage[top=1in, bottom=1in, left=1in, right=1in]{geometry}

\usepackage{utopia}
\fontfamily{utopia}

\usepackage{amsmath,amssymb,mathtools}
\usepackage{amsthm}
\usepackage{bm}
\usepackage{bbm}
\usepackage{longtable}
\usepackage{booktabs}
\usepackage{xspace}
\usepackage{setspace}
\usepackage{enumitem}
\usepackage{csquotes}
\usepackage{tikz}
\usepackage{float}
\usepackage{scalerel}
\usepackage{thmtools}

\usepackage[usenames,dvipsnames,svgnames,table]{xcolor}
\definecolor{CombinatoricaAqua}{HTML}{00698C}
\definecolor{CombinatoricaBlue}{HTML}{3A3293}
\definecolor{CombinatoricaBrown}{HTML}{66220C}
\definecolor{CombinatoricaRed}{HTML}{DF2A27}
\definecolor{HarvardCrimson}{rgb}{0.6471, 0.1098, 0.1882}

\usepackage[backref=page]{hyperref}
\hypersetup{
	unicode,
	pdfencoding=auto,
	pdfauthor={},
	pdftitle={\thetitle},
	pdfsubject={},
	pdfkeywords={},
	colorlinks=true,
	citecolor=CombinatoricaBlue,
	linkcolor=CombinatoricaAqua,
	anchorcolor=CombinatoricaBrown,
	urlcolor=HarvardCrimson}

\usepackage[justification=centering]{caption}

\makeatletter
\let\reftagform@=\tagform@
\def\tagform@#1{\maketag@@@
	{(\ignorespaces\textcolor{black}{#1}\unskip\@@italiccorr)}}
\renewcommand{\eqref}[1]{\textup{\reftagform@{\ref{#1}}}}
\makeatother

\newlist{steps}{enumerate}{1}
\setlist[steps, 1]{wide=0pt, leftmargin=\parindent, label=Step \arabic*:, font=\bfseries}

\usepackage[capitalize]{cleveref}
\crefname{claim}{Claim}{Claims}
\Crefname{claim}{Claim}{Claims}
\crefname{subclaim}{Subclaim}{Subclaims}
\Crefname{subclaim}{Subclaim}{Subclaims}
\crefname{app-corollary}{Corollary}{Corollaries}
\Crefname{app-corollary}{Corollary}{Corollaries}
\crefname{app-definition}{Definition}{Definitions}
\Crefname{app-definition}{Definition}{Definitions}
\crefname{equation}{}{}
\Crefname{equation}{}{}
\crefname{figure}{Figure}{Figures}
\Crefname{figure}{Figure}{Figures}
\crefname{lemma}{Lemma}{Lemmata}
\Crefname{lemma}{Lemma}{Lemmata}
\crefname{app-lemma}{Lemma}{Lemmata}
\Crefname{app-lemma}{Lemma}{Lemmata}
\crefname{app-proposition}{Proposition}{Propositions}
\Crefname{app-proposition}{Proposition}{Propositions}
\crefname{app-theorem}{Theorem}{Theorems}
\Crefname{app-theorem}{Theorem}{Theorems}

\declaretheoremstyle[
spaceabove=\topsep, spacebelow=\topsep,
headfont=\color{CombinatoricaBlue}\normalfont\bfseries,
bodyfont=\itshape,
]{thm}

\declaretheoremstyle[
spaceabove=\topsep, spacebelow=\topsep,
headfont=\color{CombinatoricaBlue}\normalfont\bfseries,
bodyfont=\normalfont,
]{dfn}

\declaretheoremstyle[
spaceabove=0.5\topsep, spacebelow=0.5\topsep,
headfont=\color{CombinatoricaBlue}\normalfont\bfseries,
bodyfont=\normalfont,
]{rmk}

\declaretheoremstyle[
spaceabove=0.5\topsep, spacebelow=0.5\topsep,
headfont=\color{CombinatoricaBrown}\normalfont\bfseries,
bodyfont=\normalfont,
]{assumption}

\declaretheorem[style=thm]{theorem}

\declaretheorem[style=thm]{proposition}

\declaretheorem[style=thm]{observation}

\declaretheorem[style=dfn]{definition}

\usepackage[numbers,square]{natbib}

\nc{\structurefont}[1]{\ensuremath{#1}}
\nc\term{\textit}
\nc\card[1]{\left\vert#1\right\vert}

\nc{\audz}{\ensuremath{\mathsf{0}}\xspace}
\nc{\audo}{\ensuremath{\mathsf{1}}\xspace}
\nc{\agents}[0]{\structurefont{I}}
\nc{\cost}[0]{\structurefont{c}}
\nc{\costa}[0]{\cost_{\structurefont{au}}}
\nc{\costs}[0]{\cost_{\structurefont{st}}}
\nc{\costread}[0]{\cost_{\structurefont{r}}}
\nc{\rwd}[0]{\structurefont{r}}
\nc{\rwds}[0]{\rwd_{\structurefont{st}}}
\nc{\rwda}[0]{\rwd_{\structurefont{au}}}
\nc{\s}[0]{\structurefont{t}}
\nc{\sa}[0]{\s_{\structurefont{au}}}
\nc{\sst}[0]{\s_{\structurefont{st}}}
\nc{\prob}[0]{\structurefont{p}}
\nc{\proba}[0]{\prob_{\structurefont{au}}}
\nc{\probs}[0]{\structurefont{f}_{\structurefont{st}}}
\nc{\score}[0]{\structurefont{\mathit{\gamma}}}
\nc{\coal}[0]{\structurefont{J}}
\nc{\error}[0]{\structurefont{\epsilon}}
\nc{\store}[0]{\structurefont{s}}
\nc{\report}[0]{\structurefont{\rho}}
\nc{\reports}[0]{\structurefont{\rho}}
\nc{\audit}[0]{\structurefont{e}}
\nc{\audits}[0]{\structurefont{e}}
\nc{\strat}[0]{\structurefont{\sigma}}
\nc{\stratd}[0]{\strat^{\structurefont{\circ}}}
\nc{\strath}[0]{\strat^{\structurefont{\bullet}}}

\begin{document}

\maketitle

\begin{abstract}
Decentralized storage is one of the most natural applications built on blockchains and a central component of the Web3 ecosystem. Yet despite a decade of active development---from IPFS and Filecoin to more recent entrants---most of these storage protocols have received limited formal analysis of their incentive properties. Claims of incentive compatibility are sometimes made, but rarely proven. This gap matters: without well-designed incentives, a system may distribute storage but fail to truly decentralize it.

We analyze \Shelby---a storage network protocol recently proposed by Aptos Labs and Jump Crypto---and provide the first formal proof of its incentive properties. Our game-theoretic model shows that while off-chain audits alone collapse to universal shirking, \Shelby's combination of peer audits with occasional on-chain verification yields incentive compatibility under natural parameter settings. We also examine coalition behavior and outline a simple modification that strengthens the protocol's collusion-resilience.
\end{abstract}

\section{Introduction}

One of the longest-running and arguably most natural applications on blockchains is the implementation of decentralized storage networks. Beginning in the 2010s with IPFS~\cite{Benet2014IPFS}, Sia~\cite{siaWP}, Filecoin~\cite{filecoinWP}, Storj~\cite{storjWP}, and Arweave~\cite{arweaveYP}, these systems pioneered blockchain-based approaches to content addressing and incentivized storage. More recently, the ecosystem has expanded with networks such as Crust~\cite{Crust2020Whitepaper}, Celestia~\cite{celestiaWP}, EigenDA~\cite{eigendaSpec} and Walrus~\cite{danezis2025walrus}, among others. These systems make it possible to leverage a dispersed and flexibly extensible network of service providers to store data, rewarding them for doing so. However, there remains a lack of formal incentive analysis for most of these protocols.

While \textit{distributing} storage across many nodes is an important systems property, \textit{decentralizing} storage---in the sense of aligning incentives so that independent participants reliably store their assigned data---remains a critical challenge. This distinction is often blurred: much of the existing analysis of storage protocols remains at the distributed-systems level, proving safety and liveness under the assumption that some majority of nodes are ``naively honest.''\footnote{For a recent survey, see \cite{li2024sok}.} Such assumptions may suffice for fault tolerance, but they fail to address why rational, self-interested actors would follow the protocol in the first place. The economic issue is straightforward: distributed but centrally controlled services such as AWS or Google Cloud internalize both the gains and losses from reliable storage, whereas in a decentralized network, individual storage providers do not fully internalize these externalities---bearing at most a fraction of the system-wide consequences if they choose to shirk. Audit mechanisms appear to address this by penalizing misbehavior, but a trivial solution---auditing every action on-chain---is prohibitively expensive. To keep costs down, protocols rely heavily on off-chain audits, but this creates a second incentive problem: auditors themselves must be incentivized to perform faithful audits and to report them truthfully. This ``who audits the auditors'' dilemma makes clear that without explicitly incentive-compatible design, decentralized storage risks collapsing into little more than distributed storage run on wishful assumptions.

Recently, Aptos Labs and Jump Crypto jointly introduced \Shelby \cite{whitepaper,shelbyWebsite}, a decentralized storage protocol with a two-tiered auditing system. In \Shelby, storage providers audit one another off-chain, and a subset of those audits are occasionally checked through a more costly but perfectly-verifiable on-chain review process. This design directly addresses the ``who audits the auditors'' dilemma described above, and makes \Shelby a natural case study for formal analysis. Unlike with most existing storage protocols, which provide little more than distributed-systems--style reasoning, we are able to provide a game-theoretic characterization of \Shelby's incentive compatibility.

In our model, a network of storage providers (SPs) make independent decisions about whether to store their assigned data chunks, whether to exert effort to audit others' storage behavior, and whether to truthfully report the results of any audits they conduct. We first formalize the challenge of ``auditing the auditors,'' showing that when auditors' reports must simply be relied on, with no backstop to enforce truthful auditing, the unique pure-strategy Nash equilibrium is one in which no agent provides storage or auditing services, and yet all agents (falsely) assert that all audits were conducted and passed successfully. Following the \Shelby design, we then introduce the possibility of occasional, costly on-chain verification of SPs' off-chain audits, and show that this reverses the conclusion: so long as protocol rewards  and penalties (slashing) are calibrated correctly, using occasional on-chain audits is sufficient to render the entire protocol incentive compatible. We furthermore demonstrate a sense in which these positive incentives are robust to collusion among groups of SPs. Finally, we use our framework to show that requiring SPs to submit vector commitments over audit data (in addition to their audit reports) would further strengthen \Shelby's collusion-resilience.

The key economic insight underlying these results is that infrequent but perfectly verifiable audits are sufficient to enforce truthful auditing off-chain; this in turn means that failures of storage provision will be detected and properly penalized, which makes the storage process incentive compatible as well. This mechanism---theusing costly ground-truth checks to align incentives around lower-cost interim audits---may apply more broadly in other decentralized infrastructure contexts.

\subsection{Our Contributions}

In summary, our contributions are fourfold:
\begin{itemize}
	\item We introduce a non-cooperative game theory model of service provision in a decentralized storage network with cross-provider auditing. 
	\item We use this model to formalize the ``auditing the auditors'' challenge, a general incentive challenge that arises in the autarky equilibria of these networks.
	\item We then provide the first formal incentive analysis of the \Shelby protocol, showing that the two-tiered auditing structure \Shelby uses fully addresses the aforementioned challenge, making it possible to achieve full incentive compatibility of both storage provision and auditing when SPs operate independently. Moreover, we show that this truthful equilibrium is robust to coalitions of up to half the SP population when coalition members do not have a way to commit to support each other in the auditing phase.
	\item Via our framework, we identify a potential improvement to the \Shelby protocol that would further reinforce its collusion-resilience: if SPs are additionally required to submit vector commitments to audit responses alongside their audit reports, the robustness to collusion carries over fully even when coalition members have commitment power.
\end{itemize}

\subsection{Related Work}
To our knowledge, our work is the first to provide a game-theoretic foundation for full incentive compatibility of a decentralized storage protocol in production.

Many have highlighted the importance of participation incentives---in the form of payments and other direct rewards---in decentralized storage networks (see, e.g., \cite{doan2022toward,wu2025degrees}). However, these analyses have typically been limited to high-level questions such as the need for fairness in reward distribution and ensuring that service providers can cover their operational costs, rather than reasoning about how to make the overall system fully incentive-aligned. 

Meanwhile, most public analyses of specific storage protocols emphasize distributed-systems properties (such as safety, liveness, and fault tolerance), stopping short of game-theoretic incentive analysis. For example, the Filecoin white paper \cite{filecoinWP} states a desired incentive-compatibility property but does not provide a formal proof; recent work establishes fault-tolerance guarantees \cite{wang2023security}, but, to our knowledge, no publicly available formal incentive compatibility argument for Filecoin exists. Walrus~\cite{danezis2025walrus} presents fault-tolerance guarantees without game-theoretic claims about network incentives. Arweave~\cite{arweaveYP} informally argues that its proof-of-access leads to a dominant strategy and that P2P rules incentivize ``pro-social actions,'' but does not furnish a formal proof. Storj~\cite{storjWP} discusses ``incentive alignment'' via audits, escrow, and reputation without formal incentive compatibility theorems. Celestia~\cite{celestiaWP,celestiaFDP} provides rigorous data-availability and sampling guarantees, but---as with the others---does not present a formal proof of incentive compatibility for storage providers. Sia~\cite{siaWP} and EigenDA~\cite{eigendaSpec} describe mechanisms (e.g., contracts, slashing, proof-of-custody) that govern incentives, yet do not provide formal game-theoretic proofs. Likewise, Crust's economy white paper~\cite{Crust2021Economy} articulates economic design claims rather than formal incentive compatibility results.

Meanwhile, other studies of enforcement through auditing mechanisms in distributed storage networks have focused on security and computational challenges (see, e.g., \cite{du2021enabling} and the references therein), rather than incentive-alignment of the auditing process.

The closest work to ours that we are aware of is the recent paper of \citet{vakilinia2023incentive}, which proposes a dominant-strategy incentive-compatible mechanism for decentralized storage under the assumption of a trusted (off-chain) auditing oracle network. Like \Shelby, the protocol of~\cite{vakilinia2023incentive} leverages infrequent costly verification to provide proper incentives; however, \cite{vakilinia2023incentive} uses those infrequent audits to confirm proper storage, whereas \Shelby uses them to confirm auditors' reports. As we show in our game-theoretic analysis, pushing the verification step to the auditing level removes the need to have a trusted storage oracle network---and because this drastically lowers the overall verification costs, it enables \Shelby to bring the backstop verification step on-chain.\footnote{An additional differentiator from \cite{vakilinia2023incentive} is that whereas the protocol of~\cite{vakilinia2023incentive} rely on the client detecting potential storage failures and requesting audits, \Shelby's randomized auditing scheme removes the need for this. Indeed, as our analysis in the sequel demonstrates, we show incentive-compatibility without any direct appeal or reference to the client.}

All of that said, there is significant game-theoretic precedent in the analysis of peer-to-peer storage and information networks (see \cite{ihle2023incentive} for a recent survey). And there has been recent research on market designs for ensuring truthful information revelation in decentralized physical infrastructure networks \cite{milionis2025incentive}.

And a number of authors have considered ways of compounding scarce audits into broader system-wide incentives (see, e.g., \cite{lundy2019allocation,estornell2023incentivizing,dai2025non}). Foundational to all of this research are the classic \citet{alchian1972production} framework that formalized the monitoring problem in team production, as well as and the more recent work of \citet{rahman2012but}, which introduced a game-theoretic incentive scheme for ``monitoring the monitor.''

\section{The \Shelby Protocol}\label{sec:Shelby}

We briefly summarize the design of the \Shelby protocol (for a detailed description, see \cite{whitepaper}), before proceeding to our formal analysis in \cref{sec:formal}. First, we describe \Shelby's system mechanics (\cref{sec:sm}), along with the associated payments it provides for service provision and its penalties for misbehavior (\cref{sec:p&p}). Then, we summarize the system's usage of erasure coding for fault-tolerance purposes (\cref{sec:era}).

\subsection{System Mechanics}\label{sec:sm}
At a high level, \Shelby \cite{whitepaper} delegates coordination and verification tasks to a trusted blockchain, while leaving the bulk of communication and auditing work to storage providers (SPs) off-chain.

\paragraph{Data Storage.} Each SP stores a set of data chunks and maintains an on-chain vector commitment representing the current state of its stored data. When a chunk is added or modified, the SP updates their commitment on-chain. Currently, \Shelby implements vector commitments~\cite{LNC08} using Merkle trees~\cite{Merkle80}: each chunk is a leaf in the Merkle structure, and the commitment is the Merkle root. Data chunks are approximately 1~MiB in size, and corresponding inclusion proofs are approximately 1~KiB.

\paragraph{Off-Chain Audits.} Public randomness, produced on-chain by the Aptos blockchain \cite{aptosWP}, serves as a shared seed visible to all SPs. At each block, this randomness is used to select chunks for auditing, with each selected chunk associated with a unique, commonly known, SP (the auditee). That SP must then provide a Merkle inclusion proof for the selected chunk to a subset of other SPs designated as auditors.\footnote{In practice, \Shelby uses selected subcommittees of auditors to audit a given auditee---but for expositional simplicity in our discussion and analysis, we assume that each auditee sends proofs to all other SPs.} 
This audit process continues over the course of an audit epoch. By the end of the epoch, each SP has acted both as an auditee and as an auditor multiple times.

In its auditor role, an SP collects the responses it received from auditees into an~$N$-element array, in which element~$i$ is a binary vector indicating the outcomes of the audits of SP~$i$. A value of~\audo represents a successful response; a value of~\audz indicates a failure or a missing response. If we let~$\probs$ denote the expected number of audits that each SP undergoes per epoch in their role as an auditee, then each audit report contains approximately $N \cdot \probs$ entries. Auditors retain the received inclusion proofs in local storage until the end of the \emph{next} epoch, as they may be required to present them for inspection during that epoch in the on-chain verification process we describe below. 

\paragraph{Audit Report Aggregation.} At the end of each audit epoch, each SP (in its role as an auditor) submits its audit report on-chain. Based on the full set of submitted reports, an audit score is computed for each auditee. Specifically, for each audit of SP~$i$, the audit is counted as passing if a strict majority (more than 50\%) of $i$'s auditors reported a~\audo. Non-votes (i.e., missing responses) are interpreted as~$\audz$s; and if an SP fails to submit its audit report, it is treated as if it had submitted a report consisting entirely of~$\audz$s.  The resulting audit score of SP~$i$, denoted~$\score_i \in  [0,1]$, is defined as the share of \term{passing} audits, after voting, out of all the audits~$i$ was assigned during the epoch. This computation is performed on-chain and is transparent to all participants.

\paragraph{On-Chain Audit Report Inspection.} In addition, each auditor's submitted report undergoes random inspection. Using fresh on-chain randomness, each~\audo  entry from the report is randomly selected for inspection with probability~$\proba$. The auditor is then required to provide the inclusion proof corresponding to that entry for on-chain verification. This cryptographic proof is verified directly from on-chain data (specifically, the vector commitment) and is not subject to majority voting of any kind. Failure to provide a valid proof in a timely manner is considered a failed inspection.

Finally, any SP~$i$ with an audit score~$\score_i < 1$ is subjected to additional random audits performed directly on-chain. Let $\alpha_i = 1 - {\score_i}^2$. Then, $\alpha_i \cdot C_{\max}$ of~$i$'s chunks are randomly selected for direct on-chain inclusion proof verification, where~$C_{\max}$ is a system parameter establishing the maximum audit volume.

\begin{figure}[t]
\centering
\begin{tikzpicture}[
  font=\small,
  actor/.style={draw, minimum width=2.6cm, minimum height=0.8cm, fill=blue!10, align=center},
  chain/.style={draw, minimum width=3.5cm, minimum height=0.8cm, fill=gray!10, align=center},
  msg/.style={->, thick},
  opt/.style={->, thick, dotted}
]

% Actors (columns)
\node[actor] (spi) at (0,0) {Auditee \\ SP $i$};
\node[actor] (auditors) at (5,0) {Auditors \\ SP $j$, SP $k$};
\node[chain] (aptos) at (10,0) {Aptos blockchain \\ \textit{(randomness, scoring)}};

% Time markers
\foreach \x in {0,5,10} {
  \draw[gray!40] (\x,-0.5) -- (\x,-9.0);
}

% Step 1: randomness
\draw[opt] (aptos.south) ++(0,-0.5) -- +(-2.5,-0.5) node[midway,right] {};
\draw[opt] (aptos.south) ++(0,-0.5) -- +(-2.5,-0.25) node[rotate=3, midway,above] {random seed};

\draw[opt] (aptos.south) ++(0,-2.5) -- +(-2.5,-0.5) node[midway,right] {};
\draw[opt] (aptos.south) ++(0,-2.5) -- +(-2.5,-0.25) node[rotate=3, midway,above] {random seed};

% Step 2: inclusion proofs to auditors
\draw[msg] (spi.south) ++(0,-0.3) -- ++(5,-1) node[rotate=-10, midway,above] {proof(s)};
\draw[msg] (spi.south) ++(0,-1.3) -- ++(5,-1) node[rotate=-10, midway,above] {proof(s)};
\draw[msg] (spi.south) ++(0,-2.3) -- ++(5,-1) node[rotate=-10, midway,above] {proof(s)};

% Step 3: auditors submit binary reports
\draw[ dashed] (spi.south) ++(-0.3,-3.65) -- ++(10.6,0) node[midway,right] {};
\node at (11.7,-4.1) {\textit{epoch boundary}};

\draw[msg] (auditors.south) ++(0,-4) -- ++(5,-1) node[rotate=-10, midway,above] {\audz /\audo  reports};

% Step 4: on-chain scoring
\node at (10.6,-5.5) {$\score_i$};

% Step 5: audit-the-auditor
\draw[opt] (aptos.south) ++(0,-5.3) -- ++(-5,-1) node[rotate=10, midway,below] {\qquad \qquad sample $\audo$s};
\draw[msg] (auditors.south) ++(0,-6.6) -- ++(5,-1) node[rotate=-10, midway,above] {\quad inspection proofs};

% Step 6: extra audits if score < 1
\draw[opt] (aptos.south) ++(0,-5.2) -- ++(-10,-1) node[rotate=5, midway,above] {extra audits if $\score_i<1$};
\draw[msg] (spi.south) ++(0,-6.8) -- ++(10,-1) node[rotate=-5, midway,below] {extra audit proofs};

% Time axis
\node[rotate=-90] at (-1,-5) {\textbf{Time} $\rightarrow$};

\end{tikzpicture}
\caption{Sequence of events in an audit epoch. SP~$i$ is audited by a group of SPs who submit reports to the chain. The chain computes scores, inspects auditors, and may assign extra audits if needed.}
\label{fig:time-audit}
\end{figure}
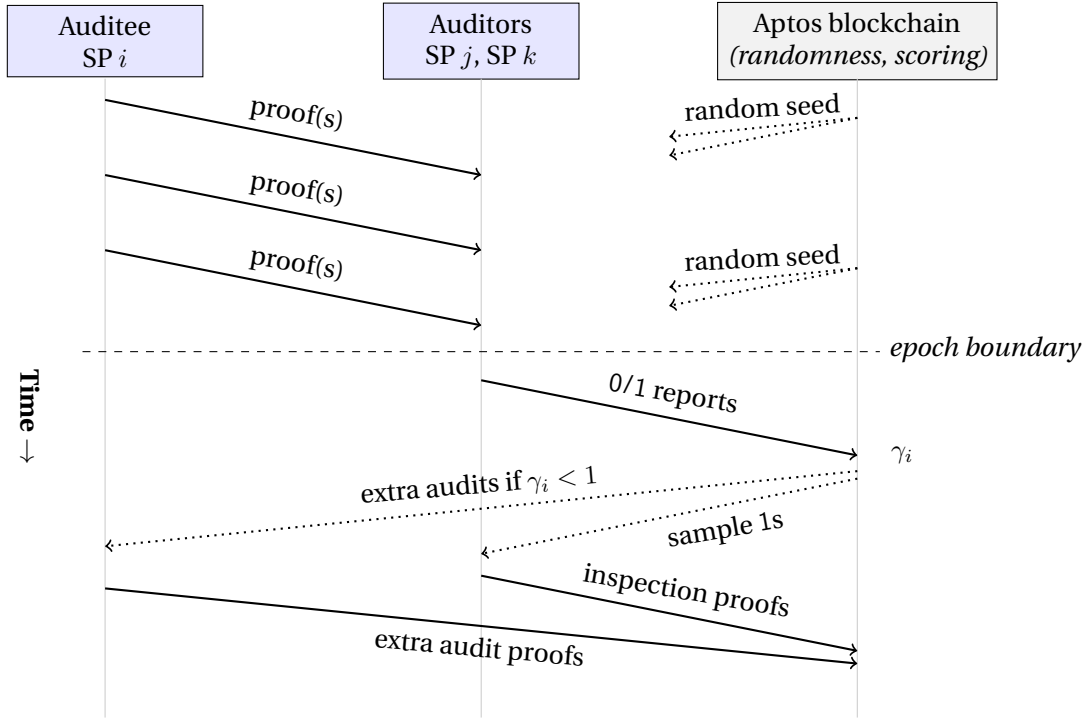

The full process flow is shown in \cref{fig:time-audit}.

\subsection{Payments and Penalties}\label{sec:p&p}

Each SP receives two types of participation rewards at the end of each epoch---one for storing its assigned data chunks, and one for (properly) performing audits:

\begin{itemize}
  \item \textbf{Storage rewards.} The storage rewards of SP~$i$ are proportional to the volume of data $i$ stores, measured in chunks, scaled by $i$'s audit score. Specifically, SP~$i$ receives $\score_i \cdot \rwds$ per stored chunk per epoch, where $\rwds$ is a fixed per-chunk storage reward.
  
  \item \textbf{Auditor rewards.} For each audit where SP~$i$ serves as an auditor, it receives a fixed reward~$\rwda$ if it reports the audit as successful (i.e., submits a \audo\ entry in the audit report).\footnote{This reward does not depend on the amount of data stored by~$i$, and is designed to encourage active participation in the protocol.} 
\end{itemize}

Misbehaving SPs may also become susceptible to slashing. All slashing events are carried out on-chain and are forensics-based. In particular, slashing cannot be imposed through voting, thus avoiding a scenario where a majority can slash a correctly-behaving minority. Slashing may occur in the following two cases:

\begin{itemize}
  \item \textbf{On-chain audit-the-auditor inspection.} If an auditor fails to provide a valid inclusion proof during a random inspection of its report, it is slashed by~$\sa$.
  \item \textbf{Additional on-chain storage audits.} If an SP with a low audit score is selected for direct on-chain audits and fails to provide a valid inclusion proof, it is slashed by~$\sst$.
\end{itemize}

\subsection{Erasure Coding and Data Reconstruction}\label{sec:era}

The \Shelby storage system employs erasure coding~\cite{roth2006introduction} to achieve efficient fault-tolerance. Very roughly, a user file is divided into multiple data chunks, and additional redundant chunks are generated (see~\cite{RScodes} for a canonical construction); the result is a set of chunks, each containing partial information about the original file, such that the full file can be reconstructed from any~$k$ chunks out of the total. For simplicity, we conflate this $k$-out-of-$N$ threshold with the requirement that at least two-thirds of SPs serve their assigned chunks faithfully.  The key advantage of erasure coding is that it enables redundancy with substantially lower storage overhead than full replication. Under the two-thirds threshold assumption, the overhead is approximately 50\%, compared to at least $3\times$ for full replication.\footnote{This approach is standard in modern large-scale storage systems~\cite{shen2023survey}.}

This redundancy has implications for the auditing scheme. In particular, an auditee that fails to store a chunk might attempt to reconstruct it on-the-fly during an audit in order to respond successfully. However, since each chunk is distinct, reconstruction is not as simple as requesting a copy from a single peer. Instead, the auditee must retrieve data from multiple different chunks---held by different SPs---to recreate its own. \Shelby's encoding and reconstruction scheme requires information from at least~$k$ distinct chunks to reconstruct a single chunk. Given that data in the system is read at the granularity of full chunks,\footnote{In specific cases (e.g., recovery of tens of GiBs at an SP), the system supports data formats allowing non-complete chunk access. This is primarily used for crash recovery and involves high data volumes, making it impractical for on-the-fly audits of random chunks.} this implies that on-the-fly reconstruction requires the auditee to read at least~$k$ chunks from~$k$ different SPs.

Denote by $\costread$ the cost of reading a single chunk from another SP. Then, answering an audit by reconstructing the chunk on-the-fly incurs a cost of at least~$k \cdot \costread$. By contrast, the expected cost of storing the chunk locally for the duration of an epoch is~$\costs$, during which it is audited~$\probs$ times in expectation. This yields the following:
\begin{observation}\label{obs:on-the-fly}
Whenever \begin{equation}
\probs \cdot k \cdot \costread > \costs,
\label{eq:n-otf}
\end{equation}
answering audits by storing the associated chunks strictly dominates answering them via on-the-fly reconstruction.
\end{observation}
Based on \Shelby's architectural documentation~\cite{whitepaper,shelbyWebsite} and pricing data from real-world systems~\cite{awsS3Pricing,gcpPricing,azureBlobPricing}, we have $k = 10$, $\probs > 3$, and estimate $\costs / \costread \approx 2.5$ (where $\probs$ and $\costs$ are normalized to a monthly basis). Thus, in practice, we expect the inequality~\eqref{eq:n-otf} holds, and storing chunks dominates reconstruction in audit response.

\subsection{Protocol Parameters}

\Cref{tab:spp} summarizes the key parameters of the \Shelby protocol relevant to our analysis.

\begin{table}[H]
\centering
\begin{tabular}{@{}ll@{}}
\toprule
\textbf{Parameter} & \textbf{Description} \\
\midrule
$\costs$ & Cost to store a chunk per epoch. \\
$\costa$ & Cost of auditing a storage provider. \\
$\costread$ & Cost of reading a chunk from another storage provider. \\
$\rwds$ & Storage reward per epoch. \\
$\rwda$ & Auditor reward per successful audit. \\
$\probs$ & Frequency that a stored chunk is audited per epoch. \\
$\proba$ & Probability that a \audo entry in an auditor report\\& \quad is selected for on-chain proof verification. \\
$\sst$ & Slashing penalty for on-chain storage audit failure. \\
$\sa$ & Slashing penalty for failing audit-the-auditor verification. \\
\bottomrule
\end{tabular}
\caption{\Shelby Protocol Parameters}\label{tab:spp}
\end{table}

\section{Incentive Analysis}\label{sec:formal}

Now, we introduce our formal framework for analyzing the incentives for storage provision and auditing under the \Shelby protocol. Building on the notation and terminology introduced in \cref{sec:Shelby}, we consider a population  $\agents$ of \term{storage providers} (\term{SPs}) indexed by $i=1,2,\ldots,N$. Each SP's action space consists of three components: 
\begin{enumerate}
    \item a binary decision as to whether to store their assigned data;
    \item a vector of binary choices as to whether to perform their assigned audits; and
    \item a vector of audit scores assessing the compliance of other SPs.
\end{enumerate}

Formally, we let $\strat_i = (\store_i,\audit_i,\report_i) \in \{0,1\} \times \{0,1 \}^{N-1} \times \{\audz,\audo\}^{N-1}$ denote SP $i$'s action space, where $\store_i=1$ indicates that SP $i$ stores its assigned data; $\audits_i^j=1$ indicates that SP $i$ properly conducts their assigned audit of SP $j$; and $\reports_i^j=\audo$ indicates that SP $i$ reports SP $j$ as having stored their assigned data. Note that we implicitly subsume the possibility that a storage provider may choose to store their assigned data yet nevertheless refuse to provide a storage proof to an auditor; in this case, the audit is ``failed'' from the perspective of the auditor, and so the auditor reports \audz.\footnote{Note that an SP may choose not to conduct audits (i.e., $\audits_i^j=0$) while still submitting a vector of audit scores. While the space of possible audit reports is richer in practice, the system treats any report that is not exactly equal to $1$ as equivalent to $0$. As a result, without loss of generality, we can model each audit report as a binary variable in our analysis.} 

The utility function of SP $i$ is given by:
\begin{equation}    
u_i(\strat_i,\sigma_{-i}) = \score_i \cdot  \rwds \cdot \mathbbm{1}_{\card{\{j : a_j^i = 1\}} > N/2} + \rwda \sum_{j \neq i}\reports_i^j -\store_i \cdot \costs - \costa \cdot \sum_{j \neq i} \audits_i^j, 
\end{equation} 
where $\rwds$ and $\rwda$, respectively,  denote the rewards for being reported as compliant and for submitting audit scores that are $\audo$s; while $\costs$ and $\costa$ are, respectively, the cost of storing data and the cost of conducting an individual audit. For simplicity and ease of notation, we express the storage reward payoff on a per-chunk basis; this is without loss of generality, as extending it to total stored data would not affect the analysis. We interpret $\rwda$ as the net reward for submitting a report with value $\audo$, since submitting an on-chain report incurs a cost regardless of whether an audit was actually performed. A storage provider who opts not to store the data (i.e., who sets $\store_i=0$) may still attempt to reconstruct the data from other providers at a reconstruction cost of $\costread$.\footnote{In practice, a storage provider attempting to retrieve data they did not store risks failure to do so. However, we take a conservative modeling approach and assume that such retrieval is always successful.} We assume that $\rwds > \costs$ and $\rwda > \costa$. Additionally, for notational convenience, we set $\probs = 1$.\footnote{Note that our positive results will continue to hold for any $\probs \geq 1$, as higher frequency of audits reinforces the audit mechanism's robustness.}

\subsection{The ``Who-Will-Audit-The-Auditor'' Problem}
As a benchmark, we analyze \Shelby's off-chain SP audits as an isolated game. Formalizing the ``auditing-the-auditors'' problem, we show that if the audit mechanism relies solely on these off-chain audits, then the fully dishonest strategy is the unique equilibrium.

Formally, we say that storage provider $i$ follows a \term{fully dishonest} strategy if they choose
\[
\stratd_i(\store_i,\audit_i,\report_i) = (0,0,0,\ldots,0,\audo,\audo,\ldots,\audo);
\]
i.e., the storage provider does not store data, assigns $\audo $ audit scores to all others, and refrains from conducting audits. 

\begin{proposition}\label{internal}
    With only off-chain audits, the unique pure strategy Nash equilibrium is (mutual) full dishonesty, i.e., 
    \[
    \forall i \in \agents,  \>\>\> \strat_i = \stratd_i.
    \]
\end{proposition}

The main intuition behind \cref{internal} is the problem of ``who audits the auditor?'' A mechanism that relies solely on off-chain audits by storage providers (SPs) offers no guarantee that audits are conducted truthfully. Without a way to verify or enforce audits, SPs have a dominant strategy to avoid audit costs and simply report successful outcomes ($\audo$s). 

In the fully dishonest equilibrium, SPs maximize their payoff---receiving all rewards without incurring costs. This outcome is robust to collusion, so long as no external funds enter the system. Thus, we see that off-chain auditing alone is insufficient---aligning incentives for storage requires a way of aligning incentives around audits.

\subsection{Truthful Behavior under On-Chain Auditing}

In this section, we augment the off-chain auditing scheme with a provably accurate on-chain mechanism, which targets storage providers suspected of underperformance and at the same time verifies a subset of audits to ensure correct reporting. Let $\proba$ denote the probability that an auditor who reports a value of $\audo$ is selected for on-chain verification of their audit proof, and let $\sa \geq 0$ represent the slashing penalty for failing this verification.

We begin by identifying a sufficient condition under which mutual dishonesty, as previously defined, is no longer a Nash equilibrium.

\begin{proposition}\label{not_equ}
     With completely verifiable on-chain auditing enforcing off-chain audits, if 
     \begin{equation}
		 \sa > \left(\frac{1-\proba}{\proba} \right)\rwda,
		 \label{eq:penalty_high_enough}
		 \end{equation}
     then (mutual) full dishonesty is not a Nash equilibrium.
\end{proposition}

Condition \eqref{eq:penalty_high_enough} guarantees the existence of a profitable deviation away from the dishonest equilibrium. Specifically, it implies that the slashed amount $\sa$ is sufficiently large to make a false report of $\audo $---as prescribed in the dishonest strategy---unprofitable. However, our goal is stronger than that: we wish to sustain an honest strategy as an equilibrium. 

\begin{definition}
We let $\strath_i$ denote the \term{honest strategy} of storage provider $i$ during an audit epoch, i.e., $\strath_i = (\store_i, \audit_i,\report_i)$, where $\store_i = 1$, $\audit_i^j = 1$ for all $j \in \agents\setminus\{i\}$, and $\report_i^j = \audo$ if and only if $\store_j = 1$ and storage provider $j$ has submitted a valid proof of storage to $i$.
\end{definition}

The honest strategy consists of two components that reflect the dual roles of each SP as a storage provider and an auditor. In the role of storage provider, the provider stores all assigned data chunks. As an auditor, the provider performs all assigned audits and reports their outcomes truthfully.

We assume that the network operates under low-level noise, i.e., there is a small probability $\error > 0$ that any SP may submit an inaccurate proof due to minor errors or imperfections in the network.\footnote{This assumption reflects the practical reality that occasional inaccuracies can arise from benign sources such as hardware faults, timing issues, or other forms of operational noise.} 

We now proceed to derive conditions under which the honest strategy profile constitutes the unique Nash equilibrium.

\begin{theorem}\label{honest}
    With completely verifiable on-chain auditing enforcing off-chain audits, if the following conditions holds:
    \begin{enumerate}[label=\roman*.]
        \item slashing discourages reporting without auditing, i.e., $\sa \geq \left( \frac{1-\proba}{\proba} \right) \rwda+\left( \frac{1}{\error \cdot \proba} \right) \costa$;
        \item audit rewards outweigh the cost of honest auditing, i.e.,  $\rwda \geq \left( \frac{1}{1-\error} \right) \costa$; and
        \item storage rewards exceed costs, i.e., $\rwds \geq \costs$;
    \end{enumerate}
    then, $\strath=(\strath_1,\ldots,\strath_N)$ is the unique Nash equilibrium.
\end{theorem}

Together, conditions (i)–(iii) ensure that the expected payoff from honest behavior---both in auditing and data storage---strictly exceeds any potential gains from deviation. Furthermore, under these conditions, the honest strategy constitutes the unique equilibrium. Condition (i) guarantees that the slashing penalty is large enough to deter storage providers from not carrying out audits while falsely reporting $\audo $, regardless of whether a proof of storage was received.\footnote{Condition (i) subsumes the inequality in \cref{not_equ}, which in this case guarantees that storage providers will not report $\audo $ in the absence of a valid proof.} Condition (ii) ensures that audit rewards are large enough to discourage providers from merely reporting $\audz $ without actually performing the audits. Finally, condition (iii) is a baseline participation constraint, ensuring that acting as a storage provider is, at minimum, individually profitable.

\subsection{Resilience to Coalitional Deviations}

In this section, we analyze the robustness of the honest equilibrium in the presence of potential collusion among storage providers (SPs) aiming to maximize their collective payoff. We consider two distinct cases. In the first and stronger case---the \textit{commitment} case---coalition members can commit in advance to a joint strategy from which no deviation is feasible ex-post.\footnote{This case can be viewed as the coalition acting as a single coordinated entity, which could in some contexts could in principle be implemented through a smart contract among the coalition members that enforces the agreed-upon strategy.} In the second, weaker case---the \textit{no commitment} case---coalition members can deviate from the agreed-upon strategy.

For the first case, we show that for coalitions comprising up to half of the SPs, the total gains from such collusion are limited, as we formalize below. We then proceed to analyze the second case.

\begin{definition}An action profile $\strat$ is \term{$\delta$-coalition resistant} if, for every subset $\coal$ of storage providers and for every joint deviation $\strat_{\coal}'$ of $\coal$, we have
$$\sum_{i\in J}u_i(\strat) \geq \sum_{i\in J}u_i \left(\strat_{\coal}',\strat_{-\coal} \right) - \delta.$$
\end{definition}

Our first result shows that, in the commitment case, for coalition of up to a half of SPs, the maximum gain a coalition can obtain by deviating from the honest strategy is bounded by the total audit cost incurred from internal auditing within the coalition.

\begin{theorem}\label{coal_dev}
Let $\coal \subseteq \{1,\ldots,N\}$ be a coalition of storage providers with $\card{\coal} < N/2$, and suppose the coalition has commitment power. Then, under conditions (i)-(iii), the (mutually) honest strategy is $\left(\card{\coal}^2 \cdot \costa\right)$-coalition resistant.
\end{theorem}

The main intuition behind \cref{coal_dev} is that, as long as the coalition size remains below half of the total number of SPs---and under the appropriate conditions outlined in \cref{honest}---it is always in the coalition members' best interest to fully store their assigned data. This inherently limits the potential gains from forming a coalition. Indeed, because the coalition members will be audited by a voting majority of SPs outside the coalition, the best joint deviation the coalition members can achieve is storing their data while not actively auditing each other. But when the coalitional equilibrium self-enforces the storing of data, within-coalition auditing is not actually necessary---so there is no loss in terms of equilibrium data storage.\footnote{This implies that if a coalition of SPs develops a more efficient method to enforce data storage than the protocol's auditing mechanism, they can, in fact, locally outperform the protocol in maintaining honest storage.} 

Moreover, if we additionally assume that coalition members do not have commitment power, then for any coalition $\coal$ with $\card{\coal} < N/2$, no joint deviation can strictly improve the payoffs of all its members. In this case, the honest strategy remains robust even under coordinated deviations.

\begin{definition} An action profile $\strat$ is an \term{$\ell$-strong equilibrium} if, for every subset $\coal$ of at most $\ell \geq 0$ storage providers and for every joint deviation $\strat_{\coal}'$ of $\coal$, there exists at least one storage provider $i \in \coal$ such that $u_i(\strat_{\coal}',\strat_{-\coal}) \leq u_i(\strat)$.
\end{definition}

Our next result shows that if coalition members do not have commitment power, then, for any coalition $\coal$ with $\card{\coal} < N/2$, the honest strategy is the unique $\card{\coal}$-strong equilibrium under the on-chain auditing mechanism.

\begin{theorem}\label{coal_res}
Let $\coal \subseteq \{1,\ldots,N\}$ be a coalition of storage providers with $\card{\coal} < N/2$, and suppose that members of the coalition do not have commitment power. Then, under conditions (i)-(iii), the (mutually) honest strategy is the unique $\card{\coal}$-strong Nash equilibrium.
\end{theorem}

\Cref{coal_res} shows that when coalition members do not have commitment power, the honest equilibrium is robust to coalitions of less than half of the storage providers. The intuition that underlies this result is that, without the ability to commit to providing services when required, storage providers within the coalition lack a strict incentive to submit valid proofs of storage ex post.

\section{Strengthening the Protocol}

\Cref{coal_dev,coal_res} provide formal coalition-resistance guarantees. However, they are not the strongest guarantees possible. In particular, when coalition members have commitment power, a coalition~$\coal$ can still obtain an aggregate benefit of~$\card{\coal}^2\cdot \costa$. We show that, while the current \Shelby auditing protocol has this caveat, the protocol can be improved to provide a stronger coalition-resistance guarantee, making the honest strategy an $\ell$-strong equilibrium also under the tougher conditions of full commitment among coalition members. This is the strongest notion of a coalition, where all members always act to the joint benefit of the coalition. In the case of a distributed system, it can be thought of as if all coalition members are controlled by a single entity. Beneficially, the proposed modification is minimal and preserves the structure of the original protocol. 

Recall that an SP, in its role as an auditor, submits an audit report on-chain. This report consists of binary values only---\audo for audits that received a successful response, and \audz otherwise. In addition, the auditor retains the successful audit responses for potential downstream audit-the-auditors inspection. We propose that the on-chain submission include an additional vector commitment over the set of successful audit responses.\footnote{Vector commitments are already used in the auditing protocol, so our proposal does not introduce new cryptographic mechanism overhead.} This commitment---a fixed number of bytes when using Merkle trees---and is added to the audit report; the underlying data itself is still retained locally by the auditor. The random inspection of audit reports will then be answered by both the audit response itself (the 1KiB inclusion proof of the audited chunk), \textit{and} an inclusion proof of the audit response verified against the addional embedded commitment in the report.

Next, we show that incorporating vector commitments alongside the binary on-chain audit matrix ensures coalition-resistance, even when coalition members have full commitment power.

\begin{proposition}\label{coal_res_mod}
Let $\coal \subseteq \{1, \ldots, N\}$ be a coalition of storage providers with $\card{\coal} < N/2$, and assume conditions (i)–(iii) hold. Suppose the auditing protocol employs a vector commitment scheme for the reports. Then, the honest strategy is the unique $\card{\coal}$-strong Nash equilibrium, even when coalition members have commitment power.
\end{proposition}

The inclusion of a vector commitment scheme ensures that even within a coalition, a storage provider will not report a value of $\audo$ without having actually received a valid proof of storage. This is because any attempt to generate the proof on-demand during the on-chain verification phase would invalidate the initial report, resulting in the provider being slashed. Thus, this modest modification to the audit protocol significantly strengthens its robustness to collusion.

\section{Discussion}

In this paper, we introduced a game-theoretic framework to analyze and prove incentive compatibility of the \Shelby decentralized storage protocol. Our negative incentive result for the model with just off-chain audits (\cref{internal}) highlights the importance of reasoning about incentives in these protocols: Even with a form of auditing in the system, incentive compatibility is not guaranteed without some way to enforce incentive alignment in the auditing process. Our positive results show that \Shelby's occasional on-chain audit verification suffices to align auditor incentives, which in turn renders the full storage network incentive compatible (\cref{honest,coal_dev,coal_res,coal_res_mod}).

\subsection*{Remarks and Further Applications}

Our results highlight a simple blockchain mechanism-design principle: \textbf{infrequent but perfectly verifiable checks can discipline cheap, high-frequency off-chain interactions}. In \Shelby, occasional on-chain verification makes truthful auditing a best response; once audits are truthful, storage becomes incentive compatible as well. The point is architectural: when a protocol pairs (i) low-cost peer reporting with (ii) a scarce but certain report verification channel, incentive-aligned behavior can emerge at reasonable cost.

\paragraph{Back-of-the-Envelope Calibration.}
Using engineering-informed orders of magnitude we gathered from conversations with the \Shelby Engineering Team~\cite{ShelbyEngineers} ($\rwds\approx\$30$/TB/month, 
$\costs\approx\$5$/TB/month,
$\costread\approx\$2$/TB, 
$\costa\approx\$10^{-7}$ per audit, 
$\rwda\approx\$5\times10^{-7}$ per audit, 
$\proba\approx2\times10^{-4}$ audit-the-auditor sampling, 
$\sa\gtrsim\$1000$,
$\probs\approx4$ audits/chunk/month), the qualitative takeaways are:
\begin{itemize}
    \item \textbf{Decentralized storage is clearly worth doing.} Indeed, $\rwds/\costs\approx6$, so the basic participation constraint is comfortably satisfied.
    \item \textbf{Honest auditing beats ``lazy negative'' reporting.} As $\rwda\approx5\costa$, truthful auditing pays more than lazily reporting all-$\audz$ even allowing for small proof/measurement noise.
    \item \textbf{False ``$\audo$'' is a dominated strategy.} A fake $\audo$ pays $\rwda\approx\$5\times10^{-7}$ but faces expected slashing $\proba\cdot\sa\approx\$0.2$; cheating is worse by a factor of $4\times10^{5}$. This margin remains substantial even if noise is as high as 1\%.
    \item \textbf{On-the-fly reconstruction remains uneconomic.} With $k=10$ and $\probs\approx4$, an auditee would need at least~40 chunk reads per month to mimic honest storage; $40\costread / \costs \approx 16$ make this highly unattractive in expectation.
\end{itemize}
These back-of-the-envelope checks are not ``tight bounds,'' but they show that the \Shelby design sits in a regime where the honest strategy is comfortable.

\paragraph{Incentivizing Whistle-Blowing is Complementary.}
Many protocols (including \Shelby) consider bounties for submitting forensic proofs of misbehavior. Our guarantees do not rely on bounties; we show that at least in the case of \Shelby, a whistleblower reward is not necessary for incentive alignment. That said, in practice, modest bounties can further deter collusion and speed fault detection, with standard griefing/DoS safeguards on submission.

\paragraph{Beyond Storage: DePIN and Data Availability.}
While the incentive design and analysis presented in this paper are rooted in the context of decentralized data storage, the same template applies in other contexts where work is off-chain but attestable, as in, for example, sensor readings, bandwidth/compute serving, or rollup data-availability sampling. In each case, peer reports carry most of the load, and a small budget of certain checks (cryptographic openings, ZK spot-checks, or L1 verifications) enforces honesty. The choice is not ``all on-chain~vs.\ all off-chain,'' but a calibrated split with a credible hammer.

\subsection*{Limitations and Future Work}

\paragraph{Model Scope.} One limitation of our analysis is the focus on Nash equilibrium as the solution concept. It remains an open question whether there are stronger implementation guarantees, such as dominant-strategy incentive compatibility, for auditing process in decentralized storage networks.

We also note that our model is static. To some degree, we view this as a feature, as it means we obtain within-period incentive guarantees---without the need to appeal to reputation or other dynamic incentives, or to consider the balance between rewards and time discounting. However, a dynamic setting may create a richer strategy space that allows for behavior not captured in our model.the

\paragraph{Identity and Sybil Resistance Calibrations.} Our analysis assumes each SP corresponds to a single player in the game. In deployment, incentive compatibility must compose with Sybil defenses (staking requirements in our case). Quantifying the combined margin is an open task.

\paragraph{Chain Budget and Latency.} The inspection rate $\proba$ trades off against L1 cost and settlement delay. Our calibration works for Aptos; systems evaluation mapping feasible $\proba$ to specific chains (throughput, gas, time-to-finality) would turn our qualitative guidance into deployment playbooks on those chains.

\begin{center}\smallskip
\rule{1in}{0.4pt}
\end{center}
\smallskip

Overall, we view our results as \emph{sufficient} conditions and a design blueprint: pair cheap peer audits with a small budget of certain checks, calibrate rewards and slashes so the expected value of cheating is deeply negative, and (optionally) bind audit artifacts with vector commitments to harden against collusion---then honest storage and truthful auditing emerge as the unique equilibrium.

\newpage
\appendix

\section{Proofs Omitted from the Main Text}

\subsection{Proof of Proposition \ref{internal}}

We first show that mutual full dishonesty constitutes a Nash equilibrium. When all SPs play $\stratd$, each SP $i$ receives
\[
u_i(\stratd_i,\stratd_{-i})=\rwds + (N-1)\rwda.
\]
This strategy incurs no storage or audit cost, while maximizing rewards. Therefore, no SP has an incentive to deviate.

To show uniqueness, suppose there were an alternative Nash equilibrium in which some SP $i$ assigns an audit score of $\audz$ to SP $j$, i.e., $\reports_i^j = \audz$. In that case, SP $i$ could profitably deviate by setting $\reports_i^j = \audo$, thereby gaining an additional $\rwda$ without incurring any further cost. This implies that reporting $\reports_i^j=\audo$ is a strictly dominant strategy. Therefore, in any equilibrium, all audit vectors must consist of $\audo$s: $\reports_i^j=\audz$ for all $i$ and $j$.

Given that SPs receive full audit rewards regardless of actual auditing, and that auditing incurs a cost, it must be the case that $\audits_i^j = 0$ for all $i$ and $j$ in equilibrium. Similarly, storing data incurs cost $\costs$ without affecting rewards if all audit reports are positive, implying $\store_i = 0$ for all $i$. Thus, mutual full dishonesty is the unique pure strategy Nash equilibrium in the internal audit game. \hfill \qed

\subsection{Proof of Proposition \ref{not_equ}}

Under mutual full dishonesty, each SP $i$ follows the strategy
\[
\stratd_i(\store_i,\audit_i,\report_i) = (0,0,0,\ldots,0,\audo,\audo,\ldots,\audo).
\]
The payoff for SP $i$ associated with $(\audits_i^j,\reports_i^j)=(0,\audo)$ is given by
\[
(1-\proba)\rwda - \proba\sa,
\]
while the payoff from setting $(\audits_i^j,\reports_i^j)=(0,\audz)$ is $0$. Thus, deviating to $(0,0)$ will be profitable if
\[
\sa > \left(\frac{1-\proba}{\proba} \right) \rwda.
\]
In this case, mutual dishonesty would no longer be a Nash equilibrium. \hfill \qed

\subsection{Proof of Theorem \ref{honest}}

We begin by showing that employing the honest strategy is an equilibrium. Assume that each storage provider (SP) $j$ stores the data (i.e., $\store_j=1$); we first show that no other SP $i$ finds it profitable to deviate from the honest strategy when choosing their audit and reporting actions $(\audits_i^j,\reports_i^j)$.

Let $u_i^j(\audits_i^j,\reports_i^j \,|\, \store_j=1)$ denote SP $i$'s  expected audit utility from choosing whether to audit SP $j$ and what to report, given that SP $j$ stores the data. Then
 \begin{align*}
     u_i^j(\strath_{i} \,|\, \store_j=1) &= \error\cdot 0 + (1-\error)\rwda-\costa \\
     u_i^j(0,\audo \,|\, \store_j=1) &= \error \left[ (1-\proba)\rwda - \proba\sa \right] +(1-\error)\rwda\\
     u_i^j(0,0 \,|\, \store_j=1) &= 0 \\
     u_i^j(1,\audz \,|\, \store_j=1) &= -\costa.
 \end{align*}
 To deter the deviation $(0,\audo)$, we require
\[
u_i^j(\strath_i \,|\, \store_j=1) \geq u_i^j(0,\audo \,|\, \store_j=1);
\] 
this implies
 \[
 \sa \geq \left( \frac{1-\proba}{\proba} \right) \rwda+\left( \frac{1}{\error \cdot \proba} \right) \costa,
 \]
 which is condition (i). 
 
 To deter the deviation $(0,\audz)$, i.e., not auditing and always reporting $\audz$, we require
 \[
  u_i^j(\strath_i \,|\, \store_j=1) \geq u_i^j(0,\audz \,|\, \store_j=1),
 \]
 which leads to
 \[
  \rwda \geq \left( \frac{1}{1-\error} \right) \costa,
 \]
 which is condition (ii).

 Next consider deviations where SP $i$ chooses not to store the data, i.e., $\store_i=0$. To render this a non-profitable deviation, it must be that any auditing SP $j$ truthfully reports $\audz$ if they detect an invalid or a missing proof. This will be the case if
 \[
 (1-\proba)\rwda -\proba\sa-\costa \leq -\costa \implies \sa \geq \left( \frac{1-\proba}{\proba} \right) \rwda,
 \]
 which is implied by condition (i).

 Given this setup, if $i$ deviates, then $i$ will be audited truthfully by other SPs and will not receive storage rewards, as a majority of them will report $\audz$ against it. Consequently, as long as
 \[
 \rwds-c_{s} \geq 0,
 \]
 SP $i$ has an incentive to store the data; when this condition---which is equivalent to (iii)---is satisfied along with conditions (i) and (ii), honest behavior constitutes a Nash equilibrium.

 We now establish uniqueness. For the sake of seeking a contradiction, assume that there exists an equilibrium in which some SP $i$ does not store their assigned data. By conditions (i) and (ii), no other SP will report a value of $\audo$ for~$i$, implying that $i$ will not receive their storage reward $\rwds$. Then, by condition (iii), $i$ has a strict incentive to deviate and store the data. Thus, it follows that in any equilibrium, all SPs must store the data.
 
 Applying conditions (i) and (ii) again, we then see that each SP will conduct their assigned audits, and report truthfully. Therefore, the honest strategy profile is the unique equilibrium. \hfill \qed

\subsection{Proof of Theorem \ref{coal_dev}}

Conditions (i)-(iii) ensure that SPs outside the coalition follow the honest strategy. In particular, by conditions (i) and (ii), and the reasoning in yhe proof of \cref{honest}, any SP $i \notin \coal$ will report a value of $\audo$ for another SP only if they have performed the audit and obtained a valid proof. Therefore, any SP within the coalition that does not store a data chunk will be unable to secure a majority of $\audo$ votes for that chunk, regardless of the coalition’s strategy. As a result, the storage-related payoff for any SP that does not store the data will be at most $0$, whereas storing the data yields a payoff of $\rwds - \costs$, which is positive by condition (iii). Hence, in any equilibrium, all SPs must store the data.

While storing the data, coalition members may coordinate to report $\audo$ scores for each other without actually performing audits. When these $\audo$ scores are themselves audited via the on-chain audit mechanism, the assigned SP will be required to submit a valid proof of storage. Given that coalition members have full commitment power, they can pre-commit to providing such proofs when requested. Thus, assuming that no SP within the coalition conduct audits on other coalition members, the maximum additional payoff gained will be $\card{\coal}^2 \cdot \costa$. \hfill \qed

\subsection{Proof of Theorem \ref{coal_res}}

Following the reasoning in the proof of \cref{coal_dev}, Conditions (i)–(iii) ensure that service providers (SPs) outside the coalition adhere to the honest strategy. As a result, any SP that fails to store the data receives a storage-related payoff of $0$, whereas storing the data yields a payoff of $\rwds - \costs$, which is positive by Condition (iii). Therefore, in any equilibrium, all SPs must store the data.

However, when coalition members lack commitment power, deviating from the honest strategy—by falsely reporting $\audo$ scores for one another without performing actual audits—ceases to be profitable. This is because there is no guarantee that valid proofs will be generated and submitted when required. Consequently, for $\card{\coal} < N/2$, the honest strategy constitutes the unique $\card{\coal}$-strong equilibrium in the no-commitment case. \hfill \qed

\subsection{Proof of Proposition \ref{coal_res_mod}}

In the case where coalition members lack commitment power, the result follows directly from \cref{coal_res}.

Now consider the case where coalition members have commitment power. The vector commitment scheme guarantees that any report of $\audo$ without a corresponding valid proof at the time of reporting will trigger slashing during on-chain verification. Consequently, any coordinated attempt within the coalition to submit $\audo$ scores without actually performing the audits carries the risk of being slashed. Therefore, under the vector commitment scheme, coalition members have no profitable deviation from the honest strategy. \hfill \qed 

\end{document}